\documentclass[12pt]{article}
\usepackage{epsfig}
\usepackage{axodraw}

\begin{document}
\normalsize
\title{\Large Mass and Decay Constant of $I=1/2$ Scalar Meson In QCD Sum Rule}
\author{Dong-Sheng Du$^{a,b}$, Jing-Wu Li$^b$ and Mao-Zhi Yang$^{a,b}$
 \thanks{Email address: duds@mail.ihep.ac.cn (Du), lijw@mail.ihep.ac.cn (Li),
  yangmz@mail.ihep.ac.cn (Yang)}\\
{\small $a$ CCAST (World Laboratory), P.O. Box 8730, Beijing 100080, China;}\\
{\small $b$ Institute of High Energy Physics, Chinese Academy of
Sciences,}\\
 {\small P.O. Box 918(4), Beijing 100049, P.R. China}\footnote{Mailing address}
}
\date{\empty}
\maketitle


\begin{abstract}
We calculate the mass and decay constant of $I=\frac{1}{2}$ scalar
mesons composed of quark-antiquark pairs based on QCD sum rule.
The quauk-antiquark pairs can be $s\bar{q}$ or $q\bar{s}$
($q=u,d$) in quark model, the quantum numbers of spin and orbital
angular momentum are $S=1$, $L=1$. We obtain the mass of the
ground sate in this channel is $1.410\pm 0.049$GeV.  This result
favors that $K_{0}^{\ast}(1430)$ is the lowest scalar state of
$s\bar{q}$ or $q\bar{s}$. We also predict the first excited scalar
resonance of $s\bar{q}$ is larger than 2.0 GeV.
\end{abstract}

\baselineskip 18pt
\section*{1. Introduction}
Glueball and scalar mesons should exist according to QCD and quark
model. Some scalar mesons below 2 GeV have been observed, such as,
i) for isospin $I=0,1$ states: $f_{0}(600)$ or $\sigma$,
$a_{0}(980)$, $f_{0}(980)$, $f_{0}(1370)$,
 $f_{0}(1500)$, $f_{0}(1710)$; ii) for $I=\frac{1}{2}$ states:
 $\kappa(900)$ and $K_0^*(1430)$ \cite{pdg2004,Lass,E791,ishida}.
The number of these scalar mesons exceeds the particle states
which can be accommodated in one nonet in the quark model. It is
believed that there are two nonets below and above 1 GeV
\cite{close1,klempt1}. The components of the meson states in each
nonet have not been completely determined yet. For the scalar
mesons below 1 GeV there are several interpretations. They are
interpreted as meson-meson molecular states \cite{ISGUR} or
multi-quark states $qq\bar{q}\bar{q}$ \cite{JAFFE}, etc.. However,
from the theoretical  point of view there must be quark-antiquark
SU(3) scalar nonet. Therefore it is important to determine the
masses of the ground states of $q\bar{q}$ with quantum number
$J^P=0^+$ based on QCD. For isospin $I=0,1$ states different quark
flavor may mix, and scalar $q\bar{q}$ states may also mix with
scalar glueball if they have the same quantum number of $J^{PC}$
and similar masses \cite{amsclo}-\cite{gluonmix53}. Some authors
have tried to determine the mixing angles of the glueball with
$q\bar{q}$ scalar mesons by using decay patterns of some scalar
mesons \cite{gluonmix53}-\cite{Wein}. These works imply that
glueball possibly mix with $q\bar{q}$ scalar mesons. For $I=1/2$
states, they cannot mix with glueball because they have strange
quantum number. The physical state is directly the $s\bar{q}$ and
$q\bar{s}$ bound state. Therefore the mass of the ground state of
$s\bar{q}$ or $q\bar{s}$ can be determined without necessity for
considering mixing effect.

QCD sum rule is a powerful tool to calculate hadronic
nonpertubative parameters based on QCD \cite{svz}. It has been
used to calculate the masses and decay constants of
$0^{-+},1^{-+},2^{++}$ mesons before and give satisfactory results
\cite{svz,mass1,reinders2,narison1}.

In this paper, we calculate the mass and decay constant of $I=1/2$
scalar meson with QCD sum rule. We find that it is impossible to
obtain $s\bar{q}$ scalar meson mass below 1 GeV from QCD sum rule.
The most favorable result for the mass of $s\bar{q}$ scalar meson
is $1.410\pm 0.049$GeV. Therefore, if $\kappa(900)$ is $s\bar{q}$
scalar bound state, this would be a big problem for QCD. This
problem can be solved by assuming that $\kappa(900)$ is irrelevant
to $s\bar{q}$ scalar channel, $<0|\bar{s}q|\kappa(900)>\sim 0$,
and $K_{0}^{\ast}(1430)$ is the scalar ground state of $s\bar{q}$
or $q\bar{s}$.  With this assumption, calculation based on QCD
will be consistent with experiment. Therefore, our result favors
that $K_{0}^{\ast}(1430)$ is the lowest scalar bound state of
$s\bar{q}$.

If this is correct, then from the approximate SU(3) flavor
symmetry, the masses of the other $J^P=0^+$ mesons in the scalar
nonet should be slightly above or below 1.4 GeV. This result would
imply that scalars with masses below 1 GeV are not dominated by
quark-antiquark pairs. This is consistent with the calculation of
lattice QCD which implies that a nonet of quark-qntiquark scalars
is in the range 1.2-1.6 GeV \cite{Ba97}.

The remaining part of this paper is organized as follows. In
Section 2, we briefly introduce the process to calculate the
scalar meson with QCD sum rule and get the Wilson coefficients for
the corresponding two-point scalar current correlation function.
Section 3 is devoted to numerical analysis and conclusion.

\section*{2. The method }


To calculate the mass of scalar $s\bar{q}$ or $q\bar{s}$ meson,
the two-point correlation function should be taken as

\begin{equation}\label{21}
   \Pi(q^{2})=i\int d^{4}x e^{iq\cdot x}\langle 0|T \{j(x)j^+(0)\}|0\rangle
\end{equation}
where $j(x)=\bar{s}(x)q(x)$, $j^+(0)=\bar{q}(0)s(0)$.

On one hand, the correlation function can be expressed based on
the dispersion relation in terms of hadron states

\begin{equation}\label{22}
  \Pi^{h}(q^{2})=\frac{1}{\pi}\int \frac{ds \hat{I}_{m}\Pi(s)}{s-q^{2}}
\end{equation}
where $\hat{I}_{m}\Pi(s)$ is the imaginary part of the two-point
correlation function, which can be obtained by inserting a
complete set of quantum states $\sum|n\rangle\langle n|$ into
Eq.(\ref{21}). The result is

\begin{equation}
  2\hat{I}_{m}\Pi(s) =  \sum_ {n} 2\pi
  \delta(s-m_{n}^{2})\langle 0|j(0)|n
  \rangle\langle n|j^+(0)|0\rangle
\end{equation}
For the scalar states $S$, its decay constant $f_S$ can be defined
through
 \begin{equation} \label{24}
   \langle 0|j(0)|S \rangle=m_{S}f_{S}
\end{equation}
where $m_S$ is the mass of the scalar state. Based on
Eq.(\ref{22})- Eq.(\ref{24}), and explicitly separating out the
lowest scalar state, the correlation function can be expressed as
\begin{equation}\label{26}
     \Pi^{h}(q^{2})=\frac{m_S^{2}f_S^{2}}{m_S^{2}-q^{2}}+\frac{1}{\pi}\int^
     {\infty}_{s^{0}}\frac{ds\rho^{h}(s)}{s-q^{2}}
\end{equation}
where $\rho^{h}(s)$ expresses the contribution of higher
resonances and continuum state, $s^{0}$ is the threshold of higher
resonances and continuum state.

On the other hand, the correlation function can be expanded in
terms of operator-product expansion at large negative value of
$q^2$.
\begin{eqnarray}
\Pi^{QCD}(q^{2})&=&i\int d^4x  e^{iq\cdot x}
 \langle 0|T \{j(x) j^+(0)\}|0\rangle \nonumber\\
 &=&C_{0}I +C_{3} \langle 0|\bar{\Psi}\Psi|0\rangle
    +C_{4} \langle 0|G^a_{\alpha\beta}G^{a\alpha\beta}|0\rangle
    +C_{5} \langle 0|\bar{\Psi}\sigma_{\alpha\beta}T^a G^{a\alpha\beta}\Psi|0\rangle
    \nonumber\\
  &~+&C_{6}\langle 0|
 \bar{\Psi}\Gamma \Psi \bar{\Psi}\Gamma^{\prime}\Psi|0\rangle +\cdots
 \label{conden}
\end{eqnarray}
where $C_i$, $i=0,3,4,5,6,\cdots$ are Wilson coefficients, $I$ is
the unit operator, $\bar{\Psi}\Psi$ is the local Fermion field
operator of light quarks, $G^a_{\alpha\beta}$ is gluon strength
tensor, $\Gamma$ and $\Gamma^{\prime}$ are the matrices appearing
in the procedure of calculating the Wilson coefficients.

For convenience later, we reexpress the above equation as
\begin{eqnarray}
  \Pi^{QCD}(q^2) &=& \frac{1}{\pi}\int \frac{ds
  \rho^{pert}}{s-q^2}+\rho^{nonp}_{3}+\rho^{nonp}_{4}+\rho^{nonp}_{5}+\rho^{nonp}_{6}
  +\cdots
  \label{conden2}
\end{eqnarray}
where $\rho^{nonp}_{3},\cdots , \rho^{nonp}_{6},\cdots$ are
contributions of condensates of dimension 3, 4, 5, 6,$\cdots$ in
Eq.(\ref{conden}).

Matching $\Pi^{h}(q^2)$ with $\Pi^{QCD}(q^2)$ we can get the
equation which relates mass of scalar meson with QCD parameters
and a few condensate parameters. In order to suppress the
contribution of higher resonances and that of condensate terms, we
make Borel transformation over $q^2$ in both sides of the
equation, the Borel transformation is defined as
$$\hat{B}_{\left|\frac{}{}\right.p^2,M^2}f(q^2)=
    \lim_{\small\begin{array}{ll}& n\to\infty \\
    & q^2\to -\infty  \\&-q^2/n=
    M^2  \end{array} } \frac{(-q^2)^n}{(n-1)!}\frac{\partial ^n}{\partial (q^2)^n}
    f(q^2).$$
After assuming quark-hadron duality, i.e., by assuming that the
contribution of higher resonance and continuum states can be
approximately cancelled by the perturbative integration over the
threshold $s^0$ \cite{colan}, the resulted sum rules for the mass
and decay constant of the scalar meson are
\begin{eqnarray}
  m_S &= &\sqrt{\frac{R_1}{R_2}}
 \label{mm}\\
 f_S&=&\frac{1}{m_S}\sqrt{e^{m_S^2/M^2}R_2}
  \label{ff}
\end{eqnarray}
where
\begin{eqnarray}\label{R_1}
 R_1 &=& \frac{1}{\pi}\int^{s^{0}}_{(m_1+m_2)^2}
     ds s\rho^{pert}(s)e^{-s/M^2}+M^4[\frac{\partial (M^2\hat{B}\rho^{nonp}_{3})}
     {\partial M^2}]
     +M^4[\frac{\partial(M^2\hat{B}\rho^{nonp}_{4})}{\partial M^2}]
     \nonumber \\&&
     +M^4[\frac{\partial(M^2\hat{B}\rho^{nonp}_{5})}{\partial M^2}]
     +M^4[\frac{\partial(M^2\hat{B}\rho^{nonp}_{6})}{\partial M^2}]
      \label{R1}
\end{eqnarray}
\begin{eqnarray}\label{R_2}
  R_2 &=& \frac{1}{\pi}\int^{s^{0}}_{(m_1+m_2)^2}
     ds \rho^{pert}(s)e^{-s/M^2}+M^2\hat{B}\rho^{nonp}_{3}
     +M^2\hat{B}\rho^{nonp}_{4}~~~~~~~~~~~~~~~~~~~~~~~~~
     \nonumber \\&&
      +M^2\hat{B}\rho^{nonp}_{5}
     +M^2\hat{B}\rho^{nonp}_{6}
      \label{R2}
\end{eqnarray}
where $\hat{B}\rho^{nonp}_{i}$ express Borel transformation of
$\rho^{nonp}_{i}$, M is Borel parameter, and $m_1$ and $m_2$ are
the masses of the two light quarks.

We need to calculate the Wilson coefficients in Eq.(\ref{conden})
to get the mass and decay constant of scalar meson. Collecting the
contribution of diagrams in Fig.1, we get the result of
$\hat{B}\rho^{nonp}_{i}$ which is listed in Appendix.

\begin{figure}[h]
\begin{center}
\scalebox{0.5}{\begin{picture}(180,80)(0,0)

\DashArrowArc(75,13.2)(58,39,141){200}
\DashArrowArc(75,86.8)(58,219,321){200}
\DashArrowLine(120,50)(150,50){3} \DashArrowLine(2,50)(33,50){3}
\end{picture}}
\scalebox{0.5}{\begin{picture}(180,80)(0,0)

\DashArrowArc(75,13.2)(58.5,39,141){200}
\DashArrowArc(75,86.8)(58.5,219,321){200}
\DashArrowLine(1,50)(31,50){3} \DashArrowLine(120,50)(151,50){3}
\GlueArc(75,2)(38,52.8,126){3}{6}
\end{picture}}

\scalebox{0.5}{\begin{picture}(180,80)(0,0)

\DashArrowArc(75,13.2)(58.5,39,141){200}
\DashArrowArc(75,86.8)(58.5,219,321){200}
\DashArrowLine(1,50)(31,50){3} \DashArrowLine(120,50)(151,50){3}
\GlueArc(75,98)(38,236,306){3}{6}
\end{picture}}
\scalebox{0.5}{\begin{picture}(180,80)(0,0)

\DashArrowArc(75,13.2)(58.5,39,141){200}
\DashArrowArc(75,86.8)(58.5,219,321){200}
\DashArrowLine(1,50)(31,50){3} \DashArrowLine(120,50)(151,50){3}
\Gluon(74,72)(74,26.8){3}{6} \put(-28,20){\large $(a)$}
\end{picture}}

\scalebox{0.5}{\begin{picture}(180,80)(0,0)
\DashArrowArc(75,13.2)(58.5,39,78){200}
\DashArrowArc(75,13.2)(58.5,100,142){200}
\DashArrowArc(75,86.8)(58.5,219,321){200}
\DashArrowLine(1,50)(31,50){3} \DashArrowLine(120,50)(151,50){3}
\put(83,68.2){$\times$} \put(59,68.2){$\times$}
\end{picture}}
\scalebox{0.5}{\begin{picture}(180,80)(0,0)
\DashArrowArc(75,13.2)(60,39,141){200}
\DashArrowArc(75,86.8)(60,218,260){200}
\DashArrowArc(75,86.8)(60,282,321){200}
\DashArrowLine(-2,50)(28,50){3} \DashArrowLine(121,50)(153,50){3}
\put(86,26){$\times$} \put(60,26){$\times$} \put(-28,20){\large
$(b)$}
\end{picture}}

\scalebox{0.5}{\begin{picture}(200,150)(0,0)
\DashArrowArc(-75,63.2)(58,39,141){200}
\DashArrowArc(-75,136.8)(58,219,321){200}
\DashArrowLine(-150,100)(-120,100){3}
\DashArrowLine(-33,100)(-2,100){3} \Gluon(-75,138)(-75,121){2}{3}
\Gluon(-75,62)(-75,79){2}{3}

\DashArrowArc(85,63.2)(58.5,39,141){200}
\DashArrowArc(85,136.8)(58.5,219,321){200}
\DashArrowLine(10,100)(41,100){3}
\DashArrowLine(130,100)(161,100){3}
\Gluon(69,138)(69,119){2}{3}\Gluon(101,119)(101,138){2}{3}

\DashArrowArc(245,63.2)(60,39,141){200}
\DashArrowArc(245,136.8)(60,219,321){200}
\DashArrowLine(168,100)(198,100){3}
\DashArrowLine(291,100)(323,100){3} \Gluon(229,62)(229,79){2}{3}
\Gluon(261,79)(261,62){2}{3} \put(80,38){\large $(c)$}
\end{picture}}

\scalebox{0.4}{\begin{picture}(450,100)(-100,-75)
\DashArrowArc(0,50)(80,282,321){200}
\DashArrowArc(0,50)(80,218,260){200}

\DashArrowArc(0,-50)(80,39,142){200}
\DashArrowLine(-103,0)(-62,0){3} \DashArrowLine(62,0)(103,0){3}
\put(13.8,-30.8){$\times$} \put(-18,-30.8){$\times$}
\Gluon(0,-18)(0,31){3}{6}\put(-3,-19.8){$\times$}

\DashArrowArc(225,-50)(80,39,78){200}\DashArrowArc(225,-50)(80,100,142){200}
\DashArrowArc(225,50)(80,218,322){200}
\DashArrowLine(118,0)(163,0){3} \DashArrowLine(287,0)(328,0){3}
\put(239,25){$\times$} \put(206,25){$\times$}
\Gluon(225,18)(225,-31){3}{6}\put(223.8,16){$\times$}

\put(100,-58){\Large $(d)$}
\end{picture}}

\scalebox{0.4}{\begin{picture}(450,280)(-100,-75)
\DashArrowLine(-93,150)(-62,150){3}
\DashArrowLine(62,150)(93,150){3} \ArrowLine(-18,178)(-62,150)
\ArrowLine(62,150)(18,178) \ArrowLine(-18,122)(-62,150)
\ArrowLine(62,150)(18,122) \put(14,175){$\times$}
\put(-21.8,175){$\times$} \put(13.8,120.2){$\times$}
\put(-21.8,120.2){$\times$}

\Gluon(31,132)(-31,168){3}{6}

\DashArrowLine(123,150)(163,150){3}
\DashArrowLine(287,150)(318,150){3} \ArrowLine(207,178)(163,150)
\ArrowLine(287,150)(243,178) \ArrowLine(207,122)(163,150)
\ArrowLine(287,150)(243,122) \put(239,175){$\times$}
\put(203.2,175){$\times$} \put(238.8,120.2){$\times$}
\put(203.2,120.2){$\times$} \Gluon(256,168)(194,132){3}{6}

\DashArrowLine(-93,0)(-62,0){3} \DashArrowLine(62,0)(93,0){3}
\ArrowLine(-18,28)(-62,0) \ArrowLine(62,0)(18,28)
\ArrowLine(-18,-28)(-62,0) \ArrowLine(62,0)(18,-28)
\put(14,25){$\times$} \put(-21.8,25){$\times$}
\put(13.8,-29.8){$\times$} \put(-21.8,-29.8){$\times$}

\Gluon(31,-18)(-31,-18){3}{6}

\DashArrowLine(123,0)(163,0){3}
\DashArrowLine(287,0)(318,0){3}\ArrowLine(207,28)(163,0)
\ArrowLine(287,0)(243,28) \ArrowLine(207,-28)(163,0)
\ArrowLine(287,0)(243,-28) \put(239,25){$\times$}
\put(203.2,25){$\times$} \put(238.8,-29.8){$\times$}
\put(203.2,-29.8){$\times$} \Gluon(256,18)(194,18){3}{6}

\put(100,-58){\Large $(e)$}
\end{picture}}
\caption{\small Diagrams for the contribution to Wilson
coefficients. (a): diagrams contribute to unit operator; (b):
diagrams contribute to bi-quark operators $\bar{\Psi}(x)\Psi(0)$;
(c): diagrams contribute to $G^{a}_{\mu\nu}G^{a\mu\nu}$; (d):
diagrams contribute to quark-gluon mixing $\bar{\Psi}(x) \Psi
(0)G^a_{\mu\nu}$; (e): diagrams contribute to four-quark operators
$\langle \bar{\Psi}\Psi\rangle ^2 $  } \label{wilsoncoefficient}
\end{center}
\end{figure}
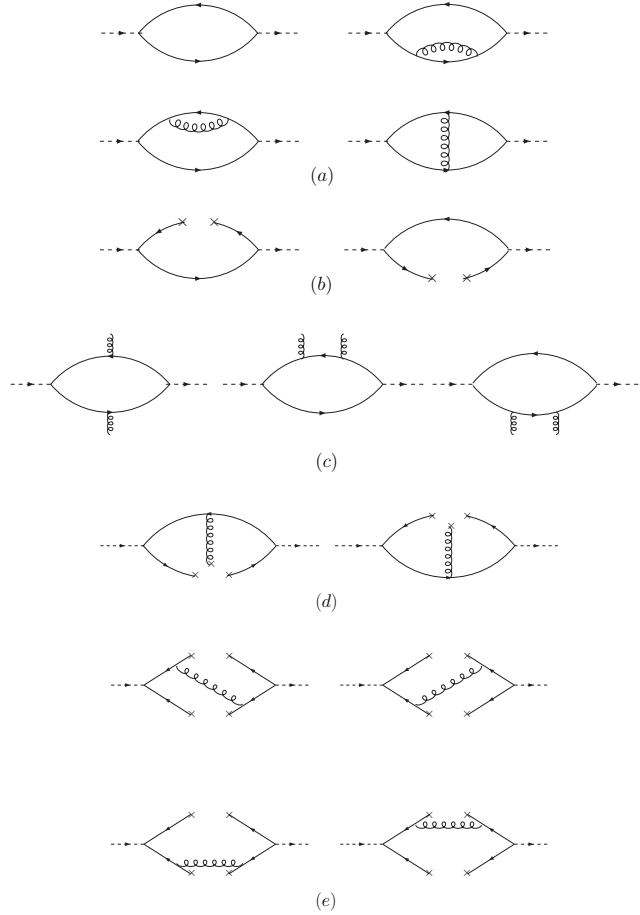

\section*{3. Numerical analysis and conclusion}
The numerical parameters used in this paper are taken as
\cite{svz,narison1}
\begin{eqnarray}
&\langle \bar{q}q\rangle =-(0.24\pm 0.01 \mathrm{GeV})^3, ~~~~~~~~
\langle \bar{s}s\rangle =m_0^2 \langle \bar{q}q\rangle
\nonumber\\[4mm]
 & \alpha_ s\langle GG\rangle=0.038\mathrm{GeV}^4,~~~~~~~~
 g\langle \bar{\Psi}\sigma T\Psi \rangle =m_0^2 \langle
\bar{\Psi}\Psi \rangle \nonumber\\[4mm]
 & \alpha_s\langle \bar{\Psi}\Psi\rangle
^2= 6.0\times10^{-5}\mathrm{GeV}^6 , m_0^2=0.8\pm 0.2
\mathrm{GeV}^2 \nonumber \\
~~~~&m_{s}=0.14\mathrm{GeV},~~~~~~~~~~~~~m_{u}\approx
m_{d}=0.005\mathrm{GeV}
\end{eqnarray}

For the choice of  Borel parameter $M^2$, as in
\cite{svz,reinders1}, we define $f_{th corr}(M^2)$ as $m(M^2)$ in
Eq.(\ref{mm}) without the continuum contribution ($s^{0}=\infty$)
and $m_{nopower}(M^2)$ as $m(M^2)$ in Eq.(\ref{mm}) without power
corrections, then define $f_{nopower}(M^2)$ as
$m(M^2)/m_{nopower}(M^2)$ and $f_{cont}$ as $m(M^2)/f_{th
corr}(M^2)$. To get reliable prediction of the mass in QCD sum
rule, $f_{cont}$ should be limited to above $90\%$ to suppress the
contribution of higher resonance and continuum, and
$f_{nopower}(M^2)$ be limited to less than $10\%$ deviation from
1, which can ensure condensate contribution much less than
perturbative contribution.

\begin{figure}[h]
 \begin{center}
 \begin{picture}(400,160)(0,0)

 \epsfig{file=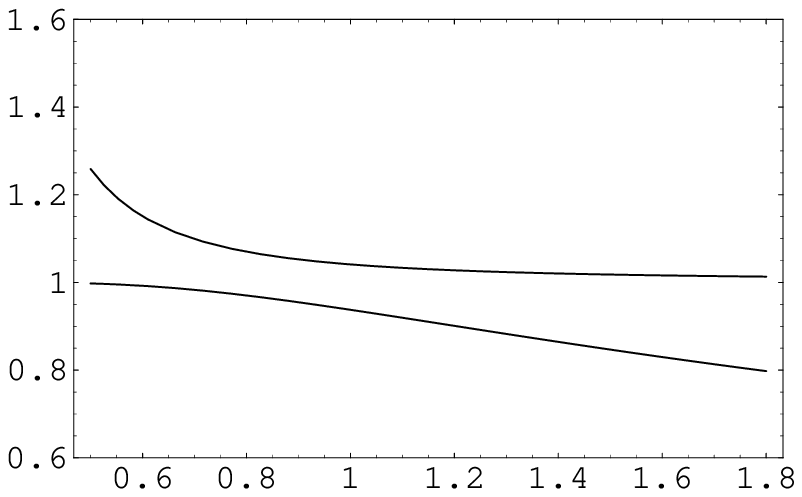,width=7.0cm,height=3.75cm}
 \epsfig{file=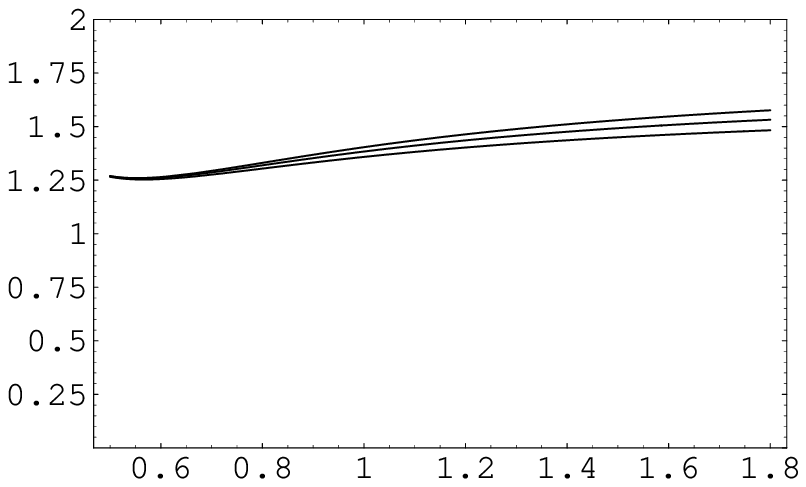, width=7.0cm,height=3.75cm}
 \put(-302,20){$A$} \put(-328,70){$B$}

\ArrowLine(-288,10)(-288,36) \ArrowLine(-313,80)(-313,50)
\put(-320,-15){$(a)$} \put(-110,-15){$(b)$}
\put(-248,-10){$M^2(\mbox{GeV}^2)$}
\put(-48,-10){$M^2(\mbox{GeV}^2)$}
 \put(-260,36){$f_{cont}$}
\put(-270,57){$f_{nopower}$}
\put(-90,85){\small{$s^0=4.8\mbox{GeV}^2$}}
\put(-90,63){\small{$s^0=4.0\mbox{GeV}^2$}}\put(-190,108){$m$\small{(GeV)}}
\end{picture}
 \end{center}
\caption{\small (a) The region between the arrow A and B is
reliable  for determining the mass ( $s^0=4.4GeV^2$). (b) The
curves correspond to the mass of scalar $s\bar{d}$ meson for the
continuum threshold $s^0=4.0GeV^2, s^0=4.4GeV^2, 4.8GeV^2$,
respectively. The central one is for $s^0=4.4GeV^2$.}
\label{figmass}
\end{figure}

There are two low mass scalar meson states with isospin $I=1/2$
and strange number $|S|=1$ found in experiment. They are
$\kappa(900)$ with mass $m_{\kappa}$ about $800\sim 900
\mathrm{MeV}$ \cite{Lass,E791,ishida}, and $K_0^{\ast}(1430)$ with
mass $m( K_{0}^{\ast}(1430))=1.412\pm 0.006\mathrm{GeV}$
\cite{pdg2004}. In theory, taking appropriate value for the
threshold parameter $s^0$, one can separate out the contribution
of the lowest resonance in QCD sum rule. We vary the value of the
threshold parameter $s^0$, and find that it is impossible to
obtain the mass of $\kappa(900)$ with the sum rule in
Eq.(\ref{mm}). There is no stable `window' for the Borel parameter
in this mass region. Therefore, if $\kappa(900)$ is the lowest
scalar state in the $s\bar{q}$ channel, it would be a big problem
for QCD sum rule. However, if we increase the value of $s^0$,
i.e., for $s^{0}=4.0\sim 4.8$ $\mbox{GeV}^2$, we does find the
stable `window' for Borel parameter, which is shown in
Fig.\ref{figmass}. The resulted stable window is in the range
$1.0<M^2<1.2 ~\mathrm{GeV^2}$.

Fig.\ref{figmass}(a) shows that between the arrows A and B, both
the contributions of condensate and higher resonance are less than
10\%. So in this region, the operator product expansion is
effective, and the assumption of quark-hadron duality does not
seriously affect the numerical result, which means that QCD sum
rule can give reliable prediction in this parameter space. For
$s^{0}=4.0\sim 4.8$ $\mathrm{GeV}^2$, the mass of scalar
$s\bar{q}$ meson in QCD sum rule is
\begin{equation}
m(s\bar{q})=1.410\pm 0.049~\mathrm{GeV} \label{mass}
\end{equation}
where the error bar is estimated by the variation of Borel
parameter in the range $1.0<M^2<1.2 ~\mathrm{GeV^2}$, the
variation of $s^0$ within $4.0\sim 4.8$ $\mathrm{GeV}^2$, the
uncertainty of higher $\alpha_s$ correction for the perturbative
diagram and the condensate parameters. The variation of Borel
parameter yields $\pm 1.8\%$ uncertainty for the mass, $s^0$
yields $\pm 2.0\%$, $\alpha_s$ correction gives $\pm 2.2\%$, the
uncertainty caused by the condensate parameters is less than
$0.6\%$. All the uncertainties are added quadratically.

The energy scale for the $\alpha_s(\mu)$ correction is taken to be
$\mu=M$. In the stable window, the range of Borel parameter is
$1.0<M^2<1.2 \mathrm{GeV^2}$, therefore $\alpha_s(M)\sim 0.5$. We
checked that the contribution of the $\alpha_s$ correction at
first order is about 2.2\%, which is not large. This can be
understood because most contribution of the $\alpha_s$ correction
is cancelled between the numerator and denominator of
Eq.(\ref{mm}). We use 2.2\% to estimate the uncertainty caused by
the higher order $\alpha_s$ corrections.

On one hand, it is impossible to obtain the mass of lower scalar
state $\kappa(900)$ from QCD sum rule for $s\bar{q}$ channel. If
regard $\kappa(900)$ as $s\bar{q}$ scalar bound state, it would be
a big problem for QCD. On the other hand, QCD sum rule can give
most favorable mass which is consistent with the mass of
$K_0^{\ast}(1430)$. Therefore it is acceptable to assume that
$\kappa(900)$ is irrelevant to $s\bar{q}$ scalar bound state, and
 \begin{equation}
 <0|\bar{s}q|\kappa(900)>\sim 0
 \end{equation}
With this assumption, $K_0^{\ast}(1430)$ can be accepted as the
lowest scalar bound state of $s\bar{q}$. Then there will be no
problem between QCD and experiment.

\begin{figure}[h]
 \begin{center}
 \begin{picture} (200,120)
\epsfig{file=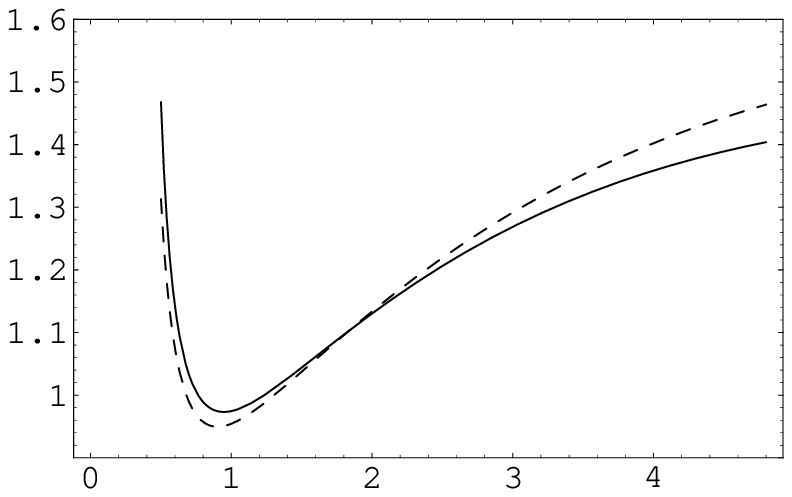,width=7.0cm,height=3.75cm}
\put(-10,-10){$s^0(\small{GeV}^2)$}
\put(-190,108){$m$\small{(GeV)}}
\end{picture}
 \end{center}
\caption{\small The possible mass result by varying the value of
the threshold $s^0$. The solid curve is for the Borel parameter
$M^2=1.0\mathrm{GeV^2}$, and the dashed one for
$M^2=1.2\mathrm{GeV^2}$. } \label{m3}
\end{figure}

One may still be afraid that there are contributions of the lower
mass state $\kappa(900)$ mixed in the result of eq.(\ref{mass}) in
fact. If this is indeed the case, the result of the sum rule may
be some weighted average of the two resonances of $\kappa(900)$
and $K_0^*(1430)$. Therefore this situation should be carefully
checked. Because the sum rule for the mass of the scalar bound
state in eqs.(\ref{mm}), (\ref{R1}) and (\ref{R2}) includes the
spectrum integration $\int_{(m_1+m_2)^2}^{s^0} ds$, in principle
one can lower the value of $s^0$ to separate the lowest bound
state. Therefore, we checked what result for the mass can be got
by lower the value of $s^0$ within the stable window $1.0<M^2<1.2
\mathrm{GeV^2}$ selected in Fig.\ref{figmass}(a). The result is
shown in Fig.\ref{m3}. It shows that for any value of $s^0$, the
possible mass is large than $960\mathrm{MeV}$,
\begin{equation}
m(s\bar{q})>960~\mathrm{MeV}
\end{equation}
Therefore the possible effect of $\kappa(900)$ can be safely ruled
out in the sum rule result in eq.(\ref{mass}). Note that the most
recent experimental result for the mass of $\kappa(900)$ from E791
collaboration is $m_{\kappa}=797\pm19\pm42 \mathrm{MeV}$
\cite{E791}.

If $K_{0}^{\ast}(1430)$ is the ground state of $s\bar{q}$ or
$q\bar{s}$, from the approximate SU(3) flavor symmetry, the masses
of the other $J^P=0^+$ mesons in the scalar nonet should be also
around 1.4 GeV. This implies that the scalars with masses less
than 1 GeV, i.e., $f_{0}(600)$, $a_{0}(980)$, $f_{0}(980)$ etc.,
can not be dominated by quark-antiquark bound states. This is
consistent with the calculation of lattice QCD which implies that
a nonet of quark-antiquark scalars is in the region 1.2-1.6 GeV
\cite{Ba97}.

Our result can be further checked by experiment. From the
threshold parameter $s^0$, we can predict that the mass of the
first excited resonance in $s\bar{q}$ scalar channel should be
larger than $\sqrt{s^0}$, that is
\begin{equation}\label{resonance}
 m^*(K_0^*)> 2.0 ~\mathrm{GeV}
\end{equation}
This prediction can be tested by experiment.

\begin{figure}[h]
 \begin{center}
 \begin{picture} (200,120)
\epsfig{file=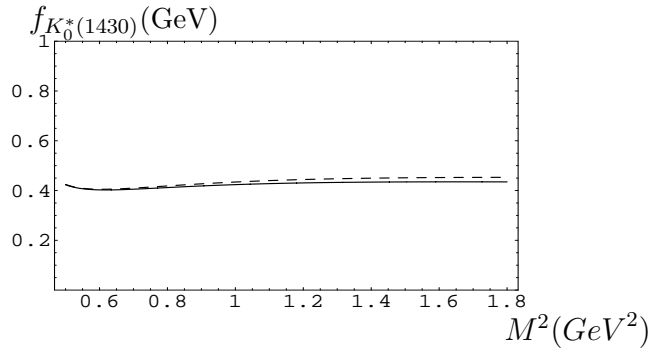,width=7.0cm,height=3.75cm}
\put(-10,-10){$M^2(\small{GeV}^2)$}
\put(-190,108){$f_{K_0^*(1430)}$\small{(GeV)}}
\end{picture}
 \end{center}
\caption{\small The decay constant of $K_0^*(1430)$ as a function
of the Borel parameter $M^2$.  The solid curve is for
$s^0=4.0\mathrm{GeV^2}$, and the dashed one for
$s^0=4.8\mathrm{GeV^2}$. } \label{f1}
\end{figure}

Next we discuss the decay constant of the two-quark scalar bound
state $s\bar{q}$. From the above analysis, we take the threshold
parameter $s^{0}=4.0\sim 4.8$ $\mathrm{GeV}^2$. Consider
$K_0^*(1430)$ as the only resonance below 2 GeV in the $s\bar{q}$
scalar channel, we can obtain the decay constant of $K_0^*(1430)$
as a function of Borel parameter $M^2$ (see eq.(\ref{ff})). The
numerical result is shown in Fig.\ref{f1}.

Fig.4 shows that the decay constant is very stable.  The
determined stable `window' is still in $1.0<M^2<1.2
\mathrm{GeV^2}$, where the continuum and condensate contribution
are restricted to be less than 15\% and 4\%, respectively. Within
this stable window, the decay constant of $K_0^{\ast}(1430)$ is
\begin{equation}\label{decay}
 f(K_0^{\ast}(1430))= 427\pm 85~\mathrm{MeV}
\end{equation}
The variation of $s^0$ yields $\pm 30\%$ uncertainty for the decay
constant, $\alpha_s$ correction gives $\pm 20\%$, the
uncertainties caused by the condensate parameters and the
variation of Borel parameter are less than $0.3\%$ and $0.1\%$,
respectively. All the uncertainties are added quadratically to
give the error bar in the above result.

Again we should check what will happen if we consider two
resonances $\kappa(900)$ and $K_0^*(1430)$ existing below 2 GeV in
our sum rule analysis. Therefore we add one more resonance into
eq.(\ref{26}), then matching $\Pi^h(q^2)$ with $\Pi^{QCD}(q^2)$ in
eq.(\ref{conden2}). By assuming quark-hadron duality to cancel the
contribution of higher resonance and continuum above 2 GeV, and
making Borel transformation in both sides, we get the Borel
improved matching equation
\begin{equation}
m_{S1}^2f_{S1}^2e^{-m_{S1}^2/M^2}+m_{S2}^2f_{S2}^2e^{-m_{S2}^2/M^2}
=R_2 \label{mR1}
\end{equation}
where $R_2$ has been given in eq.(\ref{R_2}), and $m_{S1}$,
$m_{S2}$ are fixed to be the masses of $\kappa(900)$ and
$K_0^*(1430)$, $m_{S1}=900\mathrm{MeV}$,
$m_{S2}=1410\mathrm{MeV}$. $f_{S1}$ and $f_{S2}$ are the decay
constants of the relevant scalar mesons.

\begin{figure}[h]
 \begin{center}
 \begin{picture}(400,120)(0,0)

 \epsfig{file=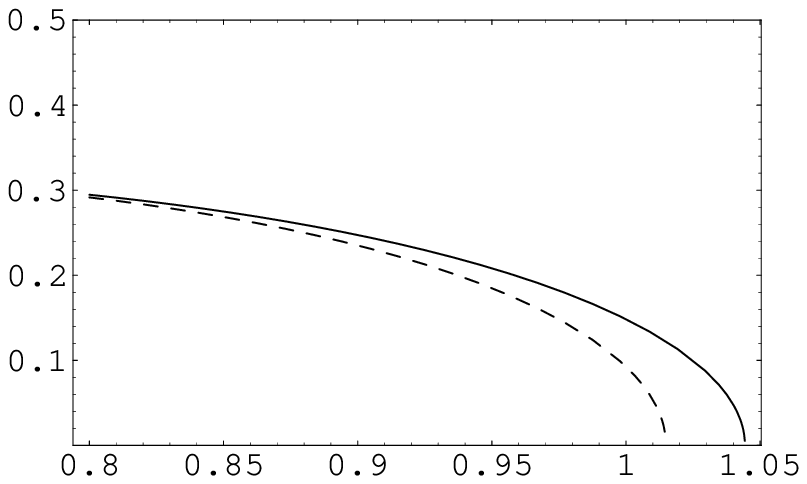,width=7.0cm,height=3.75cm}
 \epsfig{file=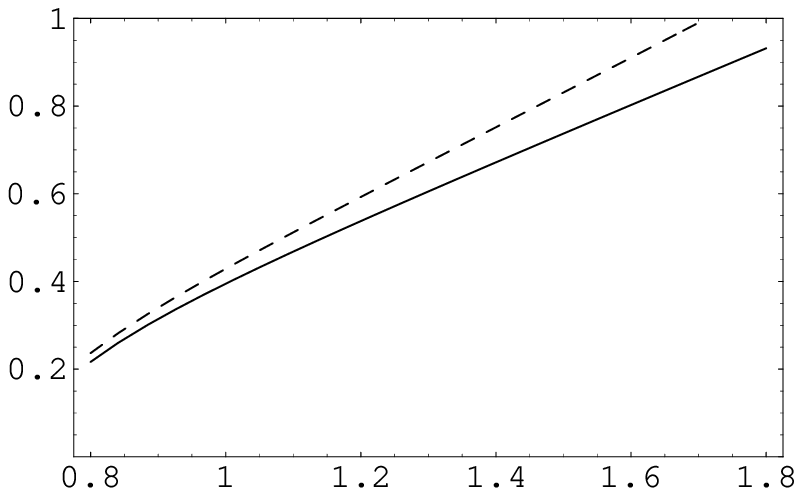, width=7.0cm,height=3.75cm}

\put(-320,-15){$(a)$} \put(-110,-15){$(b)$}
\put(-248,-10){$M^2(\mbox{GeV}^2)$}
\put(-48,-10){$M^2(\mbox{GeV}^2)$}
\put(-400,108){$f_{S1}$\small{(GeV)}}\put(-190,108){$f_{S2}$\small{(GeV)}}
\end{picture}
 \end{center}
\caption{\small The decay constants in two-resonance ansatz below
2GeV . The solid curve is for $s^0=4.0\mathrm{GeV^2}$, and the
dashed one for $s^0=4.8\mathrm{GeV^2}$. (a) The decay constant of
the low resonance $\kappa(900)$. (b) The decay constant of the
higher resonance $K_0^*(1430)$} \label{f2}
\end{figure}

Differentiate both sides of eq.(\ref{mR1}) with the operator
$d/dM^2$, we can get another equation
\begin{equation}
m_{S1}^4f_{S1}^2e^{-m_{S1}^2/M^2}+m_{S2}^4f_{S2}^2e^{-m_{S2}^2/M^2}
=R_1 \label{mR2}
\end{equation}
where $R_1$ is defined in eq.(\ref{R_1}). With eqs.(\ref{mR1}) and
(\ref{mR2}), we can obtain
\begin{eqnarray}
f_{S1}^2&=&\frac{e^{m_{S1}^2/M^2}}{m_{S1}^2(m_{S2}^2-m_{S1}^2)}(m_{S2}^2R_2-R_1)\\
f_{S2}^2&=&\frac{e^{m_{S2}^2/M^2}}{m_{S2}^2(m_{S1}^2-m_{S2}^2)}(m_{S1}^2R_2-R_1)
\end{eqnarray}
From the above result we can perform the numerical analysis for
the decay constants in the two-resonance ansatz. The numerical
result is shown in Fig.\ref{f2}.

From Fig.\ref{f2}, we can see that both the two decay constants
are unstable as a function of Borel parameter in the two-resonance
ansatz. Adding the lower resonance $\kappa(900)$ in the sum rule
analysis for the $s\bar{q}$ channel spoils the stability existing
in the one-resonance ansatz, which is shown in Fig.\ref{f1}. From
the requirement of numerical stability of QCD sum rule, the
numerical analysis of the decay constant does not favor to include
$\kappa(900)$ in $s\bar{q}$ scalar channel. In addition, we can
see from Fig.\ref{f2}a that the decay constant of the lower scalar
resonance $\kappa(900)$ tend to be zero at $M^2\sim 1.01$ and
$1.05$ GeV. This is consistent with the requirement that
$<0|\bar{s}q|\kappa(900)>\sim 0$ in the one-resonance ansatz,
where the stability window is located in the range $1.0<M^2<1.2$
GeV.

Therefore, both the analyses of the mass and decay constant of
$s\bar{q}$ scalar meson from QCD sum rule imply that $\kappa(900)$
is not dominated by quark-antiquark bound state, and the lowest
$s\bar{q}$ scalar bound state is $K_0^*(1430)$. The mass obtained
from QCD sum rule is
\begin{equation}
m(K_0^*(1430))=1.410\pm 0.049\mathrm{GeV}
\end{equation}
and the decay constant is
\begin{equation}\label{decay}
 f(K_0^{\ast}(1430))= 427\pm 85 \mathrm{MeV}.
\end{equation}

In summary, we calculate the mass and decay constant of scalar
meson $s\bar{q}$ in QCD sum rule. Our result favors that
$K_{0}^{\ast}(1430)$ is the ground state of $s\bar{q}$ scalar
bound state. If this is correct, it would imply that scalar mesons
below 1 GeV are not dominated by quark-antiquark pairs. We also
predict that the mass of the first excited resonance of $s\bar{q}$
scalar bound state is larger than 2.0 GeV.

 \vspace{1cm}

{\bf Acknowledgements} This work is supported in part by the
National Science Foundation of China, and by the Grant of BEPC
National Laboratory.

\vspace{2cm}
\begin{center}{\bf Appendix }\end{center}
Borel transformed coefficients of perturbative and nonperturbative
contributions $\hat{B}\rho^{nonp}_{i}$ in Eqs.(\ref{R_1}) and
(\ref{R_2}) are listed below
\begin{eqnarray}
   \rho^{pert}(s)&=& \{\frac{-3[(m_{1} + m_{2})^2 - s]
    \sqrt{(-(m_{1} - m_{2})^2 + s)
      (-(m_{1} + m_{2})^2 + s)}}{8\pi s}
    \nonumber \\
    &&
    + \frac{ 3s}{8\pi}\frac{13}{3}\frac{\alpha_ s(\mu)}{\pi}\}
   e^{-s/M^2}
\end{eqnarray}
where the term with $\alpha_s(\mu)$ is the radiative correction to
the perturbative contribution \cite{reinders2}, and the scale is
taken to be $\mu=M$.

\begin{eqnarray}
 \hat{B}\rho^{nonp}_{3}&=& [3M^{4}m_{1}m_{2}^{2} + 3M^{2}m_{1}^{2}m_{2}^{3} +
    m_{1}^{3}m_{2}^{4} + 3M^{6}(m_{1} + 2m_{2})]\nonumber \\
   &&
  \langle \bar{s}s\rangle\frac{e^{-m_{2}^{2}/M^{2}}}{6M^{8}}  +
 [3M^{4}m_{1}^{2}m_{2} + 3M^{2}m_{1}^{3}
     m_{2}^{2} + m_{1}^{4}m_{2}^{3}\nonumber\\&& +
    3M^{6}(2m_{1} + m_{2})]\langle \bar{d}d\rangle\frac{e^{-m_{1}^2/M^2}}{6M^{8}}
\end{eqnarray}

\begin{eqnarray}
  \hat{B}\rho^{nonp}_{4} &=&4\pi\alpha_s\langle GG\rangle\{ \frac{-3(m_{1} + m_{2})^{2}}
  {256e^{((m_{1} + m_{2})^{2}/M^{2})}M^{2}m_{1}
   m_{2}\pi^{2}}
   \nonumber\\
   &&+ (3M^{4}m_{1}^{2}m_{2} +
   3M^{2}m_{1}^{3}m_{2}^{2} + m_{1}^{4}m_{2}^{3} +
   3M^{6}(2m_{1} + m_{2}))\nonumber\\
   &&\frac{1}{288e^{(m_{1}^{2}/M^{2})}M^{8}m_{2}\pi^{2}}
  \nonumber\\
   && +
 (3M^{4}m_{1}m_{2}^{2} + 3M^{2}m_{1}^{2}m_{2}^{3} +
   m_{1}^{3}m_{2}^{4} + 3M^{6}(m_{1} + 2m_{2})) \nonumber\\
   &&\frac{1}{288e^{(m_{2}^{2}/M^{2})}M^{8}m_{1}\pi^{2}}
  \nonumber\\
   && +
 \frac{-12m_{1}(m_{1} - m_{2})^{2}m_{2} +
   M^{2}(-7m_{1}^{2 }+ 26m_{1}m_{2} -
     7m_{2}^{2})}{768
   e^{((m_{1} - m_{2})^{2}/M^{2})}M^{4}m_{1}m_{2}
   \pi^{2}}\ \nonumber\\
   && \int^{\infty}_{(m_{1}+m_{2})^{2}}dt\{\frac{3(m_{1} + m_{2})^{4}}
   {128e^{[(m_{1} + m_{2})^{2}/M^{2}]}M^{4}
    \pi^{2}(m_{1}^{2} + 2m_{1}m_{2} + m_{2}^{2 }-
     t)} \nonumber \\
   &&
    + \frac{m_{1}m_{2}(m_{1}^{2} - m_{1}m_{2} +
     m_{2}^{2} - t)t^{2}}{8e^{(t/M^2)}
    M^{4}\pi^{2}(m_{1}^{2} - 2m_{1}m_{2} +
      m_{2}^{2} - t)^{2}(m_{1}^{2 }+ 2m_{1}m_{2} +
     m_{2}^{2} - t)} \nonumber \\
   &&-
  \{(m_{1} - m_{2})^{2}[4m_{1}(m_{1} - m_{2})^{2}m_{2}
      (m_{1}^{2} - 2m_{1}m_{2} + m_{2}^{2} - t) \nonumber \\
   &&+
     M^{2}(3m_{1}^{4} - 16m_{1}^{3}m_{2} +
       26m_{1}^{2}m_{2}^{2} - 16m_{1}m_{2}^{3} +
       3m_{2}^{4} - \nonumber \\
   &&3m_{1}^{2}t +
       14m_{1}m_{2}t - 3m_{2}^{2}t)]\}\nonumber \\
   &&\frac{1}{128e^{((m_{1} - m_{2})^{2}/M^{2})}M^{6}
    \pi^{2}(m_{1}^{2 }- 2m_{1}m_{2} + m_{2}^{2} -
      t)^{2}}
   \}\nonumber \\
   &&\frac{1}{\sqrt{[(-(m_{1} - m_{2})^{2} + t)
   (-(m_{1} + m_{2})^{2} + t)]}}
 \}
\end{eqnarray}

\begin{eqnarray}
  \hat{B}\rho^{nonp}_{5} &=&g\langle \bar{\Psi}\sigma T\Psi
  \rangle\{ -\frac{m_{1}[-6M^{4} + m_{1}^{3}m_{2} +
     3M^{2}m_{1}(m_{1} + m_{2})]}{12e^{(m_{1}^2/M^2)}M^{8}}~~~~~~~
  \nonumber\\
   && -
 \frac{m_{2}[-6M^{4} + m_{1}m_{2}^{3} +
    3M^{2}m_{2}(m_{1} + m_{2})]}{12e^{(m_{2}^2/M^2)}M^{8}}
   \}
\end{eqnarray}

\begin{eqnarray}
  \hat{B}\rho^{nonp}_{6} &=& 4\pi\alpha_s\langle \bar{\Psi}\Psi\rangle
^2\{\frac{4(m_{1} + m_{2})^{2}}{9M^{2}m_{1}^{2}m_{2}^{2}} +
 [-(m_{1}^{2}m_{2}^{6})  + m_{2}^{8} \nonumber\\
   && +
   36M^{6}m_{1}(m_{1} + 2m_{2})+
   84M^{4}m_{2}^{2}(m_{1}^{2} - m_{2}^{2})\nonumber\\
   && +
   15M^{2}m_{2}^{4}(m_{1}^{2} - m_{2}^{2})]\frac{1}{81e^{(m_{2}^2/M^2)}M^{8}m_{2}^{2}
   (-m_{1}^{2} + m_{2}^{2})}\nonumber\\
   &&
 +[36M^{6}m_{2}(2m_{1} + m_{2}) +
   m_{1}^{6}(m_{1}^{2} - m_{2}^{2})\nonumber\\
   && -
   84M^{4}(m_{1}^{4} - m_{1}^{2}m_{2}^{2}) -
   15M^{2}(m_{1}^{6} - m_{1}^{4}m_{2}^{2})]\nonumber\\
   && \frac{1}{81e^{(m_{1}^2/M^2)}M^{8}m_{1}^{2}
   (m_{1}^{2} - m_{2}^{2})}
  \}
\end{eqnarray}
where $m_{1}=m_{s}$, $m_{2}=m_{q}$.

\end{document}